\documentclass[namedreferences]{solarphysics}
\usepackage[optionalrh,natbib]{spr-sola-addons}

\usepackage{graphicx,hyperref}
\usepackage{subeqn}

\usepackage[usenames]{color}

\def\cT{{\cal T}} 
\def\cK{{\cal K}} 
\newcommand{\ex}{{\mbox{\boldmath$\hat{e}_x$}}}
\newcommand{\ey}{{\mbox{\boldmath$\hat{e}_y$}}}
\newcommand{\id}{ {\rm d} }
\newcommand{\ii}{ {\mathrm i} }
\newcommand{\e}{ {\mathrm e} }
\newcommand{\bvec}[1]{ \mbox{\boldmath$#1$} }
\newcommand{\bvecs}[1]{{\scriptsize \mbox{\boldmath$#1$} }}

\begin{document}
\begin{article}
\begin{opening}

\title{Multi-Channel Three-Dimensional SOLA Inversion for Local Helioseismology}

\author{J.~\surname{Jackiewicz}$^{1,2}$ \sep
A.C.~\surname{Birch}$^{3}$\sep
L.~\surname{Gizon}$^{4,1}$\sep
S.M.~\surname{Hanasoge}$^{1,5}$\sep
T.~\surname{Hohage}$^{6}$\sep
J.-B.~\surname{Ruffio}$^{7,1}$\sep
M.~\surname{\v{S}vanda}$^{1,*}$
}

\date{Received:   / Accepted:}

\runningauthor{J. Jackiewicz \textit{et al.}}

\runningtitle{Multi-channel 3D SOLA Inversion for Local Helioseismology}

\institute{
  $^{1}$ Max-Planck-Institut f\"{u}r Sonnensystemforschung, 37191 Katlenburg-Lindau, Germany.
  \\
  $^{2}$ New Mexico State University, Department of Astronomy, 1320 Frenger Mall, Las Cruces, NM 88003, USA.
  \\
  $^{3}$ Colorado Research Associates Division, NorthWest Research Associates Inc., 3380 Mitchell Lane, Boulder, CO  80301, USA.
  \\
  $^{4}$ Institut f\"{u}r Astrophysik, Georg-August-Universit\"{a}t G\"{o}ttingen, 37077 G\"{o}ttingen, Germany.
  email: \url{gizon@astro.physik.uni-goettingen.de} \\
  $^{5}$ Department of Geosciences, Princeton University, Princeton, NJ 08544, USA.\\
 $^{6}$ Institut f\"{u}r Numerische und Angewandte Mathematik, Universit\"{a}t G\"{o}ttingen,  Lotzestrasse 16-18, 37083 G\"{o}ttingen, Germany.\\
  $^{7}$ Universit\'e de Toulouse, ISAE-Supaero, 10 avenue Edouard Belin, 31 055, Toulouse Cedex 4, France.\\
  $^{*}$ On leave from Astronomical Institute, Academy of Sciences of the Czech Republic and Faculty of Mathematics and Physics, Charles University in Prague.
  }

\begin{abstract}
Inversions for local helioseismology are an important and necessary step for obtaining three-dimensional maps of various physical quantities in the solar interior. Frequently, the full inverse problems that one would like to solve prove intractable because of computational constraints. Due to the enormous seismic data sets that already exist and those forthcoming, this is a problem that needs to be addressed. To this end, we present a very efficient linear inversion algorithm for local helioseismology. It is based on a subtractive optimally localized averaging (SOLA) scheme in the Fourier domain, utilizing the horizontal-translation invariance of the sensitivity kernels. In Fourier space the problem decouples into many small problems, one for each horizontal wave vector. This multi-channel SOLA method is demonstrated for an example problem in time--distance helioseismology that is small enough to be solved both in real and Fourier space. We find that both approaches are successful in solving the inverse problem.  However, the multi-channel SOLA algorithm is much faster and can easily be parallelized.
\end{abstract}

\end{opening}

%
%
\section{Introduction}

The inverse problem of local helioseismology is to use measurements (\textit{e.g.}, wave travel-time shifts) to infer the physical conditions in the solar interior. For  a recent review of local helioseismology, see, \textit{e.g.}, \cite{gizon2010}.  Inversions have been used to study flows and wave-speed perturbations around sunspots \citep[\textit{e.g.}][]{kosovichev1996b,gizon2000,jensen2001,zhao2001,couvidat2004,gizon2009}, flows associated with the supergranulation \citep[\textit{e.g.}][]{kosovichev1997,woodard2007,jackiewicz2008}, and global-scale flows \citep[\textit{e.g.}][]{basu1999,haber2002,zhao2004,hernandez2008}.

Essentially all inversions that have been employed in local helioseismology are linear.  These inversions are based on the assumption of a linear relationship between  perturbations to a reference model for the solar interior and the corresponding  changes in the helioseismic measurements.  The assumption of linearity is reasonable for inversions in the quiet Sun \citep[\textit{e.g.}][]{jackiewicz2007,couvidat2009}.  Within the context of linear inversions, the two main approaches are optimally localized averages \citep[OLA:][]{backus1968} and regularized least squares \citep[RLS:][]{paige1982}.  In OLA methods, the goal is to produce spatially localized estimates of conditions in the solar interior while also controlling the associated random noise.  In RLS methods, the goal is to produce a model of the solar interior that provides the best fit to the data under particular smoothness conditions.

The first three-dimensional (3D) inversions in local helioseismology were based on the RLS formalism \citep{kosovichev1996b,couvidat2005} and carried out  using the LSQR algorithm \citep[an iterative method,][]{paige1982}.  RLS corresponds to Tikhonov regularization \citep{tikhonov1963} in the mathematical literature. This approach continues to be used extensively in time--distance helioseismology \citep[\textit{e.g.}][]{zhao2001,zhao2004,zhao2007}. 
\citet{jacobsen1999} introduced the Multi-Channel Deconvolution (MCD) approach to solving the RLS equations.  In MCD, the (assumed) horizontal-translation invariance of the kernels is exploited to decouple the full RLS problem into a set of easily solvable problems, one for each horizontal wave vector. This method has been used by, \textit{e.g.}, \citet{jensen1998,jensen2001,couvidat2004}.

The 3D-OLA approach is computationally impractical for typical time--dis\-tan\-ce inversions due to the size of the matrices involved, as we will show in Section~\ref{sec:real}. An improved OLA variant, termed Subtractive OLA (SOLA) and introduced by \citet{pijpers1992}, allows one to perform fewer matrix inverse computations, yet does not reduce the sizes of the matrices. SOLA corresponds to what is known as the method of approximate inverse in the mathematical literature on regularization of inverse problems \citep{louis1990,schuster2007}. In problems where the kernel functions are separable as products of functions of horizontal position and functions of depth, it is possible to reduce the 3D-SOLA problem to a set of 2D-SOLA problems followed by 1D depth inversions \citep{jackiewicz2007a,jackiewicz2008}. However, this separation is not always possible.

In this article we will show an efficient Fourier method for carrying out  3D-SOLA problems for local helioseismology.
This method requires horizontal-translation invariance of the reference model for the solar interior and is based on an MCD approach. We will show that, unlike direct solution of the SOLA equations, this method is computationally feasible. We will also demonstrate that the multichannel approach is many orders faster than the real-space method with example computations.


\section{Setup of the Inverse Problem}
\label{sec:prob}

The solar interior is filled with numerous scatterers such as flows, magnetic fields, hot and cold spots, density and pressure anomalies, \textit{etc}. The scattering mechanism for each of these perturbations is physically different. Consider $P$ such perturbations acting on the wave field. Not accounting for the magnetic field, the thermodynamic and flow perturbations would sum to \textit{P}\,=\,5 independent quantities. For example, the three components of flow velocity, sound speed, and first adiabatic exponent.

In time--distance helioseismology \citep{duvall1993}, the measurements consist of travel times between points and concentric annuli or between points and quadrants. These travel-time measurements are performed for different choices of Fourier filters \citep[\textit{e.g.}][]{gizon2005,jackiewicz2008b}. Taken together, we assume that we have $M$ such measurements, each denoted by the running index $a$, where $1 \le a \le M$. In this article we are concerned with the local helioseismology of a small patch of the Sun near disk center, and we make the approximation that sphericity can be ignored.
Thus we adopt a Cartesian coordinate system:
\begin{equation}
  \bvec{x}=(\bvec{r},z)=(x,y,z),
\end{equation}
where $\bvec{r}$ is the horizontal position vector on the solar surface and $z$ is height.

The statement that a number of scatterers  acts on the wave field to create small shifts in travel times [$\delta\tau^a(\bvec{r})$] may be expressed  as follows \citep[\textit{e.g.}][]{gizon2002}:
\begin{equation}
  \delta\tau^a(\bvec{r})=\int_{\odot}\id^2 \bvec{r}'\id z\sum_{\beta=1}^{P} K_\beta^a(\bvec{r}'-\bvec{r},z)\delta q_\beta(\bvec{r}',z)+n^a(\bvec{r}), 
  \label{tt}
\end{equation}
where $\delta q_\beta(\bvec{r},z)$ represents the $P$ perturbations in the various physical quantities that describe the solar interior, indexed by $\beta$.
The sensitivity of a travel-time measurement [$\delta\tau^a(\bvec{r})$] to a localised change [$\delta q_\beta(\bvec{r'},z)$] is given by the travel-time sensitivity kernel [$K_\beta^a(\bvec{r'-\bvec{r}},z)$].
For point-to-annulus or point-to-quadrant measurements [$\delta\tau^a(\bvec{r})$] the position vector $\bvec{r}$ usually denotes the center of the
annulus (although there is some freedom in this convention).  
Note that in Equation (\ref{tt}) we have explicitly assumed that the background solar model as well as the sensitivity kernels are invariant under horizontal translations. Sensitivity kernels result from a forward modeling under the single-scattering Born approximation \citep{gizon2002}. The integral is taken over the volume of the Sun. The term $n^a(\bvec{r})$ is the stochastic noise of the travel-time measurement [$\delta\tau^a(\bvec{r})$] due to the forcing of waves by turbulent convection. The travel-time noise covariance matrix [${\mathbf \Lambda}$] has elements
\begin{equation}
\Lambda_{ab}(\bvec{r}_i-\bvec{r}_j)={\rm Cov}\left[n^a(\bvec{r}_i),n^b(\bvec{r}_j)\right].
\label{ttnoise}
\end{equation}
Details about the computation of ${\mathbf \Lambda}$ can be found in \citet{gizon2004}.

The general OLA inversion problem for time--distance helioseismology seeks to find an estimate of $\delta q_\alpha$ at any chosen target position [$\bvec{r}_0;z_0$], given a set of $\delta\tau$, $K$, and $\Lambda$. In other words, we are looking for a linear combination of the travel times such that
\begin{equation}
\delta q_\alpha^{\rm inv}(\bvec{r}_0;z_0)=\sum\limits_{i=1}^N \sum\limits_{a=1}^M w_a^\alpha(\bvec{r}_i-\bvec{r}_0;z_0)\delta\tau^a(\bvec{r}_i),
\label{invpb}
\end{equation}
is an estimate of $\delta q_\alpha(\bvec{r}_0;z_0)$. The weights $w_a^\alpha(\bvec{r}_i-\bvec{r}_0;z_0)$ are the unknowns of the problem. In the above equation, $N=(2n+1)^2$ is the total number of horizontal position vectors [$\bvec{r}_i$]. Throughout the article we assume that the travel times are given on a uniform square grid with sampling $h_x$ in both horizontal directions and with $n_x=n_y=2n+1$ pixels on each side.


\section{Subtractive Optimally Localised Averaging}
\label{sectionsola}

Using the Equations (\ref{tt}) and (\ref{invpb}) we have:
\begin{eqnarray}
\delta q_\alpha^{\rm inv}(\bvec{r}_0;z_0)&=& \int_{\odot}\id^2 \bvec{r}'\id z \sum_{\beta=1}^{P} \left[ \sum\limits_{i=1}^N \sum\limits_{a=1}^M w_a^\alpha(\bvec{r}_i-\bvec{r}_0;z_0)  K_\beta^a(\bvec{r}'-\bvec{r}_i,z) \right]\delta q_\beta(\bvec{r}',z) \nonumber\\
&& + \sum\limits_{i=1}^N \sum\limits_{a=1}^M w_a^\alpha(\bvec{r}_i-\bvec{r}_0;z_0)n^a(\bvec{r}_i).\label{qinv1}
\end{eqnarray}
We can rewrite Equation~(\ref{qinv1}) as: 
\begin{subequations}
\begin{eqnarray}
\delta q_\alpha^{\rm inv}(\bvec{r}_0;z_0)&=&\int_{\odot}\id^2\bvec{r}'\id z \,\cK^\alpha_\alpha(\bvec{r}'-\bvec{r}_0,z;z_0)\delta q_\alpha(\bvec{r}',z)\label{final1}\\
&&+\int_{\odot}\id^2\bvec{r}'\id z\!\!\!\sum_{\beta=1,\beta\neq\alpha}^{P}\!\!\!\cK_\beta^\alpha(\bvec{r}'-\bvec{r}_0,z;z_0)\delta q_\beta(\bvec{r}',z) \label{final2}\\
&&+\sum\limits_{i=1}^N \sum\limits_{a=1}^M w_a^\alpha(\bvec{r}_i-\bvec{r}_0;z_0)n^a(\bvec{r}_i) \label{final3},
\end{eqnarray}
\label{final}
\end{subequations}
where the functions $\cK^\alpha_\beta$ are averaging kernels given by
\begin{equation}
\cK_\beta^\alpha(\bvec{r}, z;z_0)=\sum\limits_{i=1}^N \sum\limits_{a=1}^M w_a^\alpha(\bvec{r}_i;z_0)K_\beta^a(\bvec{r}-\bvec{r}_i,z),\quad \forall \beta\in[1,P],
\label{avereal}
\end{equation}
where $\beta$ is a running index between $1$ and $P$ that labels the physical quantities.

The term (\ref{final1}) on the right-hand side of Equation~(\ref{final}) is what we are searching for, \textit{i.e.} the quantity $\delta q_\alpha$ convolved with the averaging kernel $\cK_\alpha^\alpha$. If the averaging kernel $\cK_\alpha^\alpha(\bvec{r},z;z_0)$ is well localized in the horizontal (around $\bvec{r}=\bvec{0}$ and vertical (around $z=z_0$) directions, we will recover a  smoothed estimate of $\delta q_\alpha$. 

The term (\ref{final2}) is the leakage from the other perturbations $\beta\neq\alpha$ into $\delta q^{\rm inv}_\alpha$. Ideally, one would like all averaging kernels $\cK_\beta^\alpha$  with $\beta \ne \alpha$ to be zero. 

The term (\ref{final3}) represents the propagation of random noise from the travel times into the inverted $\delta q_\alpha^{\rm inv}$.

The SOLA method consists of searching for the inversion weights $w_a^\alpha(\bvec{r};z_0)$ so that the averaging kernels $\cK_\beta^\alpha$ resemble user-supplied target functions [$\cT^\alpha_\beta$], while keeping error magnification as small as desired. This can be achieved by minimising the cost function
\begin{subequations}
\begin{eqnarray}
{\cal X}_{\alpha}\left(\bvec{w}^\alpha; \mu\right)&=& \int_{\odot}\id^3 \bvec{x}\sum_{\beta=1}^{P}\left[\cK^\alpha_\beta(\bvec{x};z_0)-\cT^\alpha_\beta(\bvec{x};z_0)\right]^2\label{min1}\\
&+&\mu \sum\limits_{i=1}^N \sum\limits_{a=1}^M \sum\limits_{j=1}^N \sum\limits_{b=1}^M w_{a}^{\alpha}(\bvec{r}_i;z_0)\Lambda_{ab}(\bvec{r}_i-\bvec{r}_j)w_{b}^{\alpha}(\bvec{r}_j;z_0) .
\label{min2}
\end{eqnarray}
\label{min}
\end{subequations}

Equation~(\ref{min}) has two main components: The first term (\ref{min1}), the ``misfit'', is a measure of how the averaging kernels [$\cK^\alpha_\beta$] match the target functions [$\cT^\alpha_\beta$]. To minimise the cross-talk, the target has only one non-vanishing component [$\cT_\alpha^\alpha$], which we write as the product of a 2D Gaussian in the horizontal coordinates by a 1D function of the vertical coordinate. Thus we write
\begin{equation}
\cT^\alpha_\beta(\bvec{x};z_0)= {\cal C} \exp{\left(\frac{-||\bvec{r}||^2}{2s^2}\right)} f(z;z_0)\,\delta_{\beta\alpha}, \qquad \forall \beta\in[1,P],
\label{targ}
\end{equation}
where $\delta_{\beta\alpha}$ is the Kronecker $\delta$-function. Typically, the function $f$ peaks at a desired target depth $z=z_0$. The constant ${\cal C}$ is taken so that the spatial integral of $\cal{T}^\alpha_\alpha$ is unity. The parameter $s$  controls the width of the Gaussian.

The second term (\ref{min2}) of Equation~(\ref{min}) is proportional to the variance [$\sigma_\alpha^2$] of the random noise in $q_\alpha^{\rm inv}$ due to the propagation travel-time noise:
\begin{equation}
\sigma_\alpha^2 \equiv \sum\limits_{i=1}^N \sum\limits_{a=1}^M\sum\limits_{j=1}^N \sum\limits_{b=1}^Mw_{a}^{\alpha}(\bvec{r}_i;z_0)\Lambda_{ab}(\bvec{r}_i-\bvec{r}_j)w_{b}^{\alpha}(\bvec{r}_j;z_0).
\label{varnoise}
\end{equation}
The trade-off parameter $\mu$ in Equation~(\ref{min}) is chosen to provide a satisfactory trade-off between the misfit and the noise; this choice is somewhat subjective.

The minimization is also subject to the constraints
\begin{equation}
\int_{\odot} \cK_\beta^\alpha(\bvec{x};z_0)\,\id^3\bvec{x}=\delta_{\beta\alpha}, \qquad \forall\beta \in [1,P] ,
\label{constraint}
\end{equation}
which ensure that the inverted quantity is normalised appropriately.


\section{Linear System of Equations}
\label{sec:real}

The solution to the SOLA problem defined in Section~\ref{sectionsola} has been traditionally solved in real space for the one- and two-dimensional cases \citep[\textit{e.g.}][]{pijpers1994}. Below we write the problem for the 3D case.

We recast the optimization problem of Equation~(\ref{min}) subject to the constraints (\ref{constraint}) by minimising the function
\begin{equation}
{\cal L}_{\alpha}\left(\bvec{w}^\alpha, \bvec{\lambda}; \mu\right) = {\cal X}_{\alpha}\left(\bvec{w}^\alpha; \mu\right) + 2\sum_{\beta=1}^{P} \lambda^\beta\left(\int_{\odot}\cK^{\alpha}_{\beta}(\bvec{x};z_0)\id^3\bvec{x}-\delta_{\beta\alpha}\right)
\end{equation}
with respect to the vector of weights $\bvec{w}^\alpha$ and a vector of Lagrange multipliers $\bvec{\lambda}$. There are $M \times N$ unknown weights $w^\alpha_{a}(\bvec{r}_j;z_0)$ and $P$ unknown Lagrange multipliers $\lambda^\beta$. The minimization consists of solving the following set of $M\times N + P$ equations:
\begin{equation}
  \frac{\partial}{\partial w^{\alpha}_a(\bvec{r}_i; z_0)}\left\{{\cal L}_{\alpha}\right\}=0, \quad\forall (a,i)\in[1,M]\times[1,N], \label{minw}
\end{equation}
and  
\begin{equation}
\frac{\partial}{\partial \lambda^\beta}\{{\cal L}_{\alpha}\}=0, \quad\forall \beta\in[1,P]. \label{minl}
\end{equation}
Equations (\ref{minw}) and (\ref{minl}) imply
\begin{eqnarray}
  \sum\limits_{j=1}^N \sum\limits_{b=1}^M A_{ab}(\bvec{r}_i-\bvec{r}_j)w^{\alpha}_b(\bvec{r}_j;z_0)&+&\sum_{\beta=1}^{P} C_{a\beta}\lambda^\beta=t^\alpha_a(\bvec{r}_i;z_0), \nonumber\\
  && \qquad\forall\, (a,i)\in[1,M]\times[1,N],
\label{real1}
\end{eqnarray}
and
\begin{equation}
  \sum\limits_{j=1}^N \sum\limits_{b=1}^M C_{b\beta} w_b^\alpha (\bvec{r}_j;z_0)=\delta_{\beta\alpha}, \qquad \forall\beta\in[1,P],
\label{real2}
\end{equation}
where we define
\begin{eqnarray}
A_{ab}(\bvec{r}_i-\bvec{r}_j) &\equiv& \int_{\odot}\id^2 \bvec{r}\,\id z\sum_{\beta=1}^{P} K^a_\beta(\bvec{r}-\bvec{r}_i,z)K^b_\beta(\bvec{r}-\bvec{r}_j,z)\nonumber\\
&+& \mu\,\Lambda_{ab}(\bvec{r}_i-\bvec{r}_j),\label{amat}\\
C_{a\beta}&\equiv&\int_{\odot} K_\beta^a(\bvec{r}-\bvec{r}_i,z)\id^2 \bvec{r}\,\id z=\int_{\odot} K_\beta^a(\bvec{x})\,\id^3\bvec{x},\label{cmat}\\
t^\alpha_a(\bvec{r}_i;z_0) &\equiv& \int_{\odot} K^a_\alpha(\bvec{r}-\bvec{r}_i,z)\cT_{\alpha}^\alpha (\bvec{x};z_0)\,\id^3\bvec{x}.\label{tmat}
\end{eqnarray}

Equations (\ref{real1}) and (\ref{real2}) form a system of $M\times N+P$ linear equations. The weights $w^\alpha_a(\bvec{r}_i;z_0)$ can be obtained by matrix inversion, or some equivalent linear solver. Inspecting Equations~(\ref{amat}), (\ref{cmat}) and (\ref{tmat}), one sees that the dimension of the matrix that is to be inverted is $(MN+ P)\times(MN+P)$. 
An advantage of the SOLA method (see Section~\ref{sectionsola}) compared to the OLA method is that the systems of equations for different target functions differ only in the right-hand sides and not
in the matrix. Therefore the large matrix in Equation~(\ref{amat}) has to be set up and inverted only once. For example, after inversion, we can infer the physical quantity at any depth using a new $t_a^\alpha(\bvec{r}_i;z_0)$.

Let us now discuss the computational cost of the SOLA scheme  for  typical time--distance inversions. The computational costs depends very much on the size of the matrix ${\mathbf A}$ to be inverted. Typically, ${\mathbf A}$ may have over $10^{14}$ elements (over a petabyte!), which do not fit in computer memory. In practice, we do not solve the full problem, but truncate the sums over $j$ in Equations (\ref{real1}) and (\ref{real2}). We consider a restricted number of convolution ``shifts'' [$\bvec{r}_j=(x_j,y_j)$] such that 
\begin{eqnarray}
- n_{\rm shifts} h_x \le & x_j &\le n_{\rm shifts} h_x, \\
- n_{\rm shifts} h_x \le & y_j &\le n_{\rm shifts} h_x, 
\end{eqnarray}
where $h_x$ is the horizontal sampling. The total number of shifts [$N_{\rm shifts} = (2n_{\rm shifts}+1)^2$] must be much less than $N$ so that the problem can now be solved.
The minimum number of shifts that is acceptable depends on the size of the target function and the horizontal extent of the sensitivity kernels.  In the example of Section~\ref{sec:example}, we have $n_{\rm shifts} = 45$. The smaller the number of shifts, the worse the approximation of the problem and the worse the localisation and the noise of the answer.

If one takes the same number of parameters as in the (2+1)D flow inversion of \citet{jackiewicz2008} then $P=3$ (the three components of velocity),  $n_{\rm shifts}=10$, so that $N_{\rm shifts}=441$, and $M=3\times 5\times 20$ for three geometries (waves propagating in ``outward minus inward'', ``West\,--\,East'', and ``North\,--\,South'' directions), five ridges ($f,\ p_1,\ p_2,\ p_3,$ and $p_4$ modes), and twenty different radii. The problem would then require inverting a matrix of size  $\approx~\!\!\!100\,000\times 100\,000$ that occupies tens of gigabytes of memory. Furthermore, it is well-known that matrix inverse operations scale as $O({\cal N}^3)$, where ${\cal N}=(NM+P)$ is the length of the matrix on one side. Even without memory issues, this calculation becomes intractable very quickly.

Another issue that should be mentioned is the computation of the kernel-overlap integral in Equation~(\ref{amat}). This computation is extremely expensive in real space.
However it can be sped up very significantly by transforming to horizontal Fourier space, where it becomes a simple multiplication. Convolution operations are $O({\cal N}^2)$ in the real space but reduce to $O({\cal N}\ln {\cal N})$ when performed in Fourier space. In order to describe this step explicitly, let us define the discrete Fourier transform.  Any function  $f$ and its Fourier transform $\tilde{f}$ are related according to
\begin{eqnarray}
\tilde{f}(\bvec{k}) &=& \frac{h_x^2}{(2\pi)^2}\sum_{\bvecs{r}}f(\bvec{r})\,\e^{-\ii\bvecs{k}\cdot\bvecs{r}},\label{fft1}\\
f(\bvec{r}) &=& h_k^2\sum_{\bvecs{k}}\tilde{f}(\bvec{k})\,\e^{\ii\bvecs{k}\cdot\bvecs{r}},
\label{fft}
\end{eqnarray}
where $h_k=2\pi/(n_xh_x)$ is the sampling in Fourier space. The horizontal wave vector $\bvec{k}$ takes the discrete values $\bvec{k}_{pq}=ph_k\ex+qh_k\ey$, where $p$ and $q$ are integers in the range [$-n,n$]. Using this definition of the Fourier transform, we obtain
\begin{eqnarray}
h_x^2 \sum\limits_{l=1}^N \,\sum_{\beta=1}^{P} K^a_\beta(\bvec{r}_l-\bvec{r}_i,z)K^b_\beta(\bvec{r}_l-\bvec{r}_j,z)=\qquad \qquad\qquad  \qquad&&\nonumber \\
 (2\pi h_k)^2  \sum_{\bvecs{k}}\e^{\ii\bvec{k}\cdot(\bvec{r}_i-\bvec{r}_j)} \sum_{\beta=1}^{P} \tilde{K}_{\beta}^{a*}(\bvec{k},z)\tilde{K}_\beta^b(\bvec{k},z),&&
\end{eqnarray}
where we used the fact that the Fourier transform of $K_{\beta}^a(-\bvec{r})$ is $\tilde{K}_{\beta}^{a*}(\bvec{k})$ since $K_{\beta}^a$ is real.
In this form the kernel-overlap integral is computed much faster than in the Equation~(\ref{amat}). 
The SOLA inversion examples presented later were computed using these equations. In order to avoid the edge effects resulting from the implicit periodicity assumed by the Fourier transform, we padded the sensitivity kernels and noise covariance matrices with zeros over a zone as wide as the size of the widest sensitivity kernel.


\section{Solution in Fourier Space}
\label{sec:four}
In this section we fully exploit the horizontal-translation invariance of the sensitivity kernels and rewrite the entire problem in Fourier space. Using the definition of Equation~(\ref{fft1}), the Fourier transforms of  Equations~(\ref{real1}) and (\ref{real2}) are
\begin{equation}
h_k^4N\sum_{b=1}^{M}\tilde{A}_{ab}(\bvec{k})\tilde{w^{\alpha}_b}(\bvec{k};z_0)+\delta_{\bvecs{k},\bvecs{0}}\sum_{\beta=1}^{P} C_{a\beta}\lambda^\beta=h_k^2\tilde{t}^\alpha_a(\bvec{k};z_0),\qquad \forall\,a,\bvec{k}
\end{equation}
and
\begin{equation}
h_k^2N\sum_{b=1}^{M}C_{b\beta}\tilde{w}_b^{\alpha}(\bvec{0};z_0)=\delta_{\beta\alpha}, \qquad \forall\beta,
\end{equation}
where $\bvec{k}$ takes the discrete values $\bvec{k}_{pq}=ph_k\ex+qh_k\ey$, with $p$ and $q$ in the range [$-n,n$]. This set of equations can be written conveniently in matrix form as
\begin{equation}
  \left\{\begin{array}{l l}
      \mbox{if}\quad \bvec{k}\neq\bvec{0},\qquad &  h_k^2 N\tilde{\mathbf{A}}(\bvec{k})\tilde{\bvec{w}}^\alpha(\bvec{k};z_0)=\tilde{\bvec{t}}^\alpha(\bvec{k};z_0), \\\\
      \mbox{if}\quad \bvec{k}=\bvec{0},\qquad   & \left[\begin{array}{c c}
          h_k^4N\tilde{\mathbf{A}}(\bvec{0}) & \mathbf{C}\\
          \mathbf{C}^{\rm T} & \mathbf{0}
        \end{array}\right]
      \left[\begin{array}{c}
          \tilde{\bvec{w}}^\alpha(\bvec{0};z_0)\\
          \bvec{\lambda}
        \end{array}\right]
      =\left[\begin{array}{c}
          h_k^2\tilde{\bvec{t}}^\alpha(\bvec{0};z_0)\\
          \bvec{U}^\alpha/(h_k^2N_x^2)
        \end{array}\right],
    \end{array}
  \right.
\label{solfour}
\end{equation}
where the vector $\tilde{\bvec{t}}^\alpha(\bvec{k})=\left[\tilde{t}^\alpha_1(\bvec{k};z_0)\ \tilde{t}^\alpha_2(\bvec{k};z_0)\  \dots\  \tilde{t}^\alpha_M(\bvec{k};z_0)\right]^{\rm T}$ and the 
matrix $\tilde{\mathbf{A}}(\bvec{k})=\left[\tilde{A}_{ab}(\bvec{k})\right]$ have the following elements
\begin{equation}
\tilde{A}_{ab}(\bvec{k}) = (2\pi)^2\int_{z_{\rm bot}}^{z_{\rm top}}\sum_{\beta=1}^{P}\tilde{K}^{a^*}_{\beta}(\bvec{k},z)\tilde{K}^b_\beta(\bvec{k},z)\id z + \mu\tilde{\Lambda}_{ab}(\bvec{k}),\label{afour}
\end{equation}
\begin{equation}
\tilde{t}^\alpha_a(\bvec{k};z_0) = (2\pi)^2\int_{z_{\rm bot}}^{z_{\rm top}}\tilde{K}^{a*}_{\alpha}(\bvec{k},z)\tilde{\cT}_{\alpha}^\alpha(\bvec{k},z;z_0)\,\id z,
\end{equation}
where $z_{\rm bot}$ and $z_{\rm top}$ are the bottom and top heights of the computation box. Furthermore,  $\bvec{U}^\alpha=[\delta_{1,\alpha}\ \delta_{2,\alpha}\ \dots\ \delta_{P,\alpha}]^{\rm T}$, $\tilde{\bvec{w}}^\alpha(\bvec{k})=\left[\tilde{w}^\alpha_1(\bvec{k};z_0)\ \tilde{w}^\alpha_2(\bvec{k};z_0)\dots\right.$ $\left.\tilde{w}^\alpha_M(\bvec{k};z_0)\right]^{\rm T}$, and $\mathbf{C}=\left[C_{a\beta}\right]$ has elements given by Equation (\ref{cmat}).

In Fourier space the problem decouples into many small problems, one for each horizontal wave vector $\bvec{k}$. These small problems are completely independent and therefore can be solved in a parallel fashion. The solution $\tilde{w}^\alpha_a(\bvec{k})$ is constructed for each $\bvec{k}$ separately. By analogy to the RLS Multi-Channel Deconvolution \citep{jacobsen1999}, we call the current approach multi-channel SOLA or MCD SOLA.

For each wave vector, the matrix to be inverted is much smaller than in the real-space case. For each wave vector $\bvec{k} \ne \bvec{0}$, the matrix is of size $M^2$. Taking the same parameters as in Section~\ref{sec:real}, the Fourier approach would only need $441$ inversions of matrices of size 300\,$\times$\,300.  This would result in an increased speed by more that five orders of magnitude over the real-space method for this realistic example. Note that there is no need to truncate the problem anymore ($n_{\rm shifts}=n$).

We provide below expressions for the averaging kernel [Equation (\ref{avereal})] and the variance of the noise [Equation (\ref{varnoise})] in terms of the Fourier transform of the weights:
\begin{equation}
\cK_{\beta}^{\alpha}(\bvec{r},z;z_0)= h_k^4 N  \sum\limits_{\bvecs{k}}  \e^{\ii\bvecs{k}\cdot{\bvecs{r}}} \sum\limits_{a=1}^{M} \tilde{w}_{a}^{\alpha}(\bvec{k};z_0)\tilde{K}^a_\beta(\bvec{k},z) ,
\end{equation}
\begin{equation}
\sigma^2_\alpha=h_k^6N^2 \sum\limits_{a=1}^M \sum\limits_{b=1}^M\sum_{\bvecs{k}}\tilde{w}_a^{\alpha *}(\bvec{k};z_0)\tilde{\Lambda}_{ab}(\bvec{k})\tilde{w}_b^{\alpha}(\bvec{k};z_0).
\end{equation}
We emphasize that the averaging kernel is now computed as  a matrix multiplication instead of a convolution.

The inferred solar property $\delta q^{\rm inv}_\alpha$ at position $(\bvec{r},z_0)$ is 
\begin{equation}
\delta q^{\rm inv}_\alpha(\bvec{r};z_0) = N h_k^4  \sum\limits_{\bvec{k}} \e^{\ii\bvec{k}\cdot \bvec{r}} \sum\limits_{a=1}^M \tilde{w}^{\alpha*}_{a}(\bvec{k};z_0) \delta\tilde{\tau}^a(\bvec{k})\ ,
\end{equation}
where $\delta\tilde{\tau}^a(\bvec{k})$ is the Fourier transform of the travel-time maps.


\section{Example Inversion for Sound Speed}
\label{sec:example}

\subsection{Setup}

\begin{figure}
  \centerline{
    \includegraphics[width=.45\textwidth]{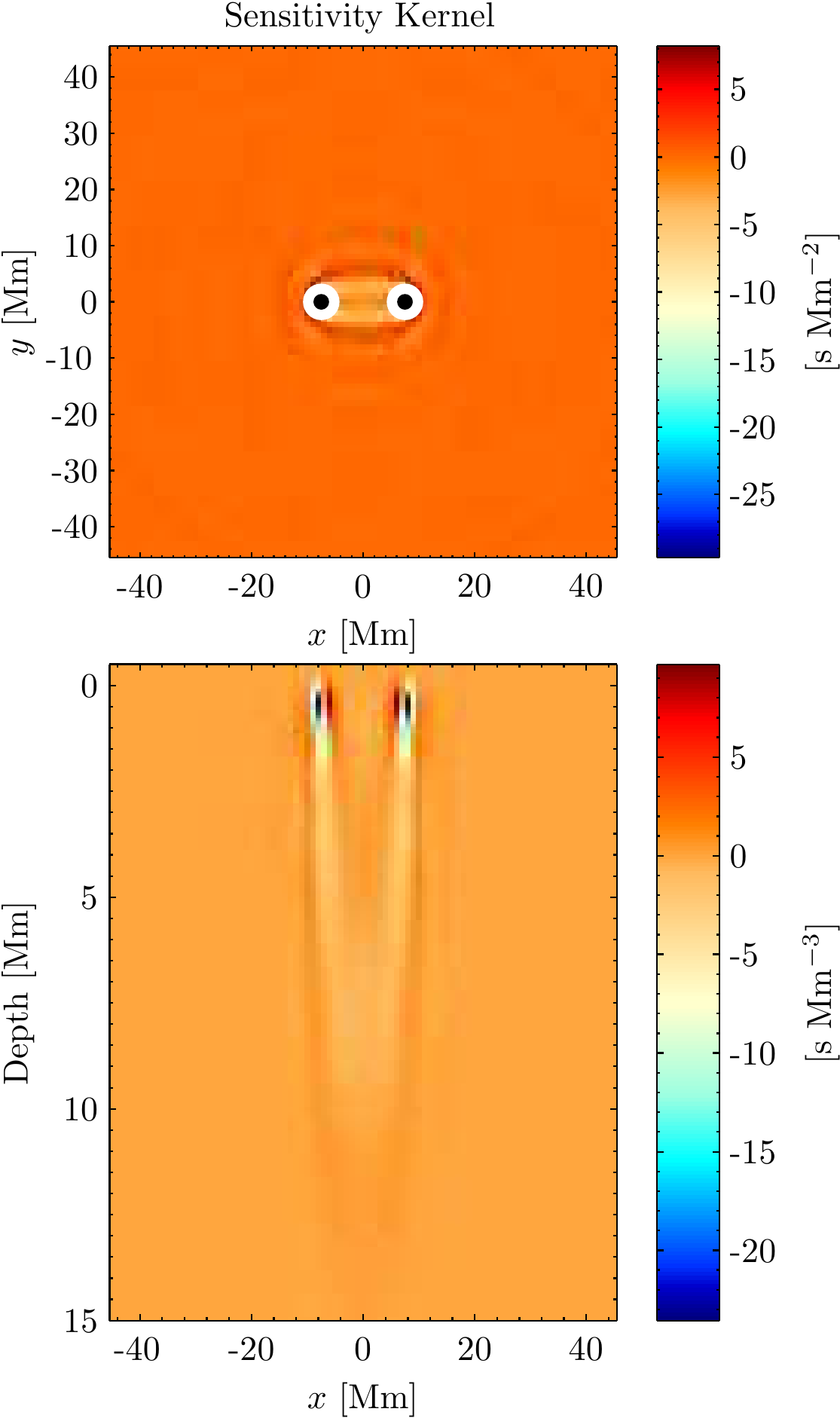}\hspace{.05\textwidth}
    \includegraphics[width=.45\textwidth]{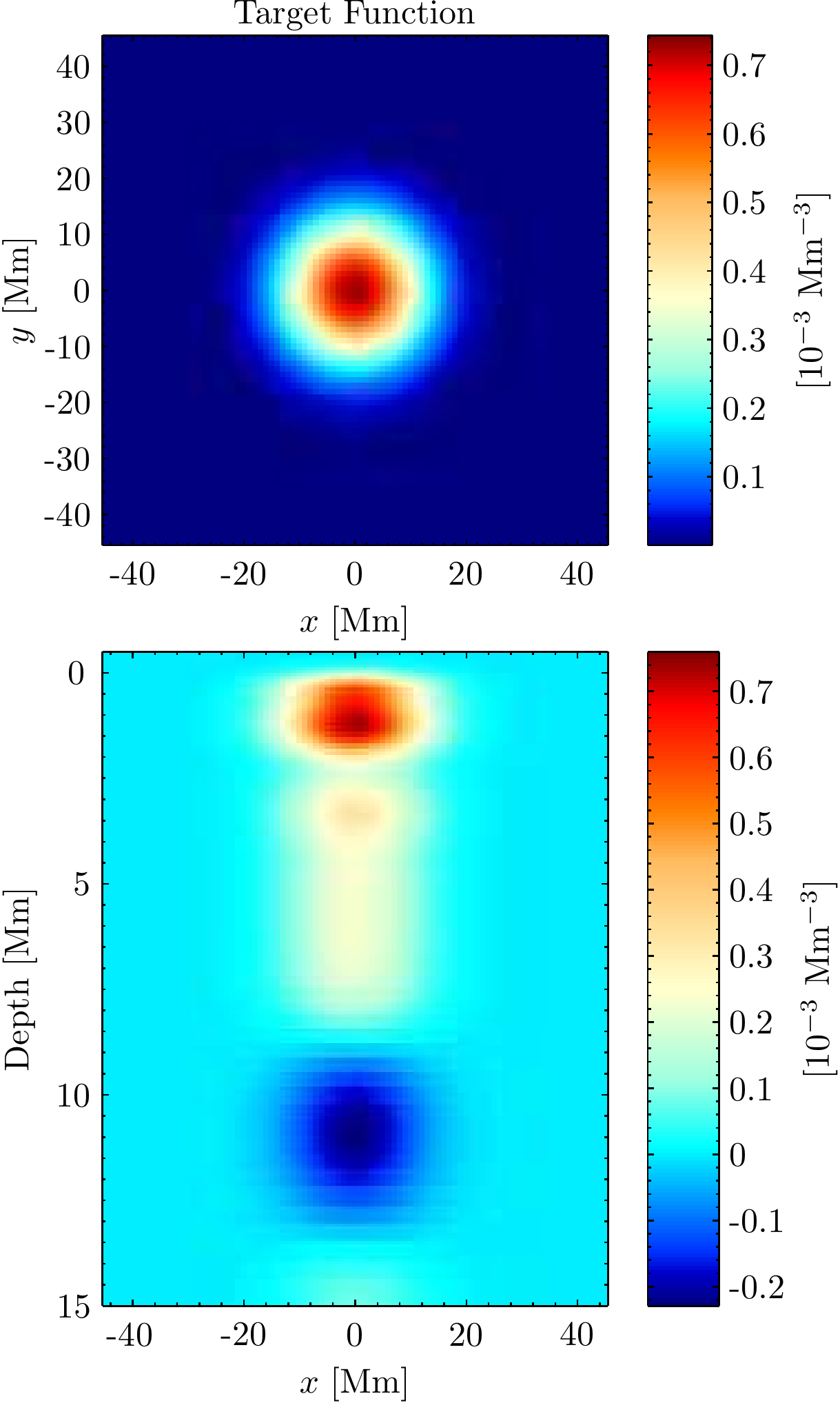}
  }
  \caption{Inputs to the example inversion. The point-to-point Born sensitivity kernel for sound speed is shown in the left column, and a target function with a full width at half maximum 20 Mm in the right column. The kernel in the $x-y$ plane (top left) is the 1D  spatial integral of the kernel over depth. The black and white circles denote the two observation points, separated by 15 Mm. The depth slice in the lower-left panel is taken along the $y=0$ line. The horizontal cut of the target function (top right) is at a depth of $1$~Mm. The  depth slice (lower right) is also along $y=0$. The depth profile of the target was computed according to Equation~(\ref{targdep}).}
  \label{senskern}
\end{figure}

We now show a rather simple example of a time--distance helioseismic inversion to demonstrate the Multichannel SOLA method and compare it to its real-space counterpart. For simplicity, we will only consider one mean (mn) point-to-point travel-time measurement with the distance between the two observation points fixed at $\Delta=15$~Mm.  For consistency with the measurement, we have computed a point-to-point Born-approximation sensitivity kernel according to \citet{birch2004a}.  This kernel gives the sensitivity of mean travel times to the sound speed perturbation [$\delta c^2/c^2$]. No prior filtering has been done, \textit{i.e.} the whole model power spectrum is used. We are not  interested in specific types of kernels for this example problem, only a comparison between inversion methods and to prove that the MCD inversion works. This sound-speed kernel, denoted $K^{\rm\scriptsize{mn}}_{c^2}$ according to the conventions in Section~\ref{sec:prob}, is shown in Figure~\ref{senskern}. This kernel has $91\times 91$ elements in the horizontal direction and 80 elements in the vertical direction.

\begin{figure}
  \centering
  \includegraphics[width=\textwidth]{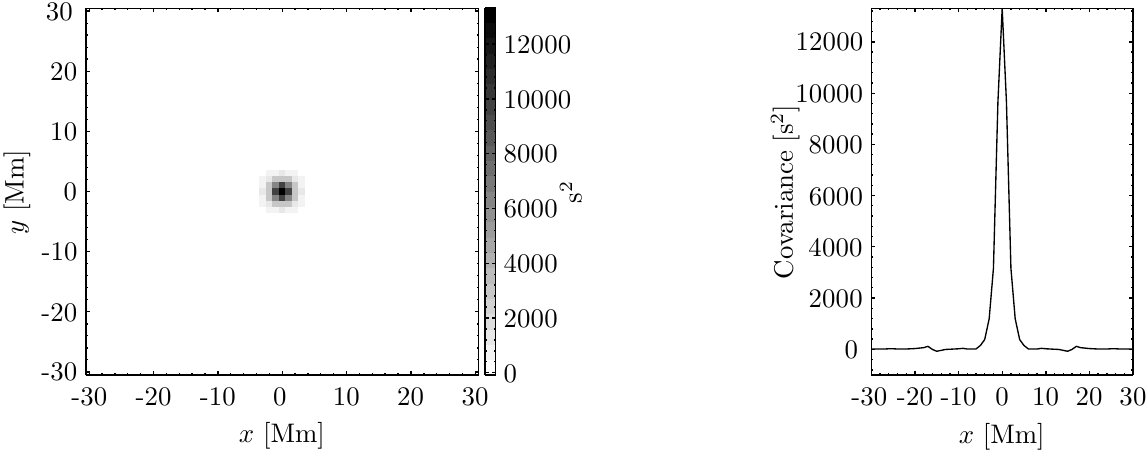}
  \caption{Model noise covariance matrix for mean point-to-point travel times with $\Delta=15$ Mm for  the example inversion. The left panel shows the noise covariance matrix in units of s$^2$ as a function of horizontal coordinates. The right panel is a cut through the matrix along the $y=0$ line. The averaging time for this noise estimation is eight hours.}
  \label{fig:noisecov}
\end{figure}

The second input quantity kept fixed for our example  inversion is the target function, shown alongside the kernel in Figure~\ref{senskern}. This 3D function has a Gaussian horizontal structure with a full width at half maximum of $20$~Mm [see Equation~(\ref{targ})]. Since we are only working with one single sensitivity kernel (one $\Delta$) in this example, it would be futile to attempt to obtain an averaging kernel peaked at some chosen $z=z_0$, since the kernel itself possesses no such depth properties. Therefore, again for simplicity, we choose a depth profile of the target function by horizontally integrating the sensitivity kernel at each depth coordinate to obtain a one-dimensional curve according to
\begin{equation}
  f(z;z_0) = \int \id^2\bvec{r}\,K_{c^2}^{\scriptsize{\rm mn}}(\bvec{r},z).
  \label{targdep}
\end{equation}
The  1D-curve $f(z; z_0)$ is then combined with the horizontal Gaussian to construct  the 3D target as in Equation~(\ref{targ}). This choice for $f(z;z_0)$ keeps the example as simple as possible. Note that the target in Figure~\ref{senskern} has a weak negative lobe beneath a depth of 10 Mm as a result of the depth profile of the sensitivity kernel.

The final input quantity to the inversion is the noise covariance matrix defined in Equation~(\ref{ttnoise}) and denoted in this case  as $\Lambda_{\rm mn,mn}$. We compute the covariance from the model power spectrum according to \citet{gizon2004} and show the results in  Figure~\ref{fig:noisecov}. What this matrix tells us is how two mean point-to-point travel-time measurements are spatially correlated due to  noise as the pairs of observation points are moved around with respect to each other. Note that for this case there is  a significant correlation only when the measurements  are made within about 10 Mm of each other.

\subsection{Comparison of the Real-Space and Fourier-Space Solutions}

We compute a set of real-space inversions and one Fourier inversion using the  input quantities. The Fourier inversion is not computed in parallel for this example. For each inversion we generate a trade-off, or ``L'' curve \citep{hansen1998} by choosing ten values of the parameter $\mu$ [see Equation~(\ref{afour})]. The values of $\mu$ are chosen to span the space of misfit and noise. One trade-off curve is generated for the Fourier inversion, but several are generated for the real-space inversion, each corresponding to a different number of shifts employed. The possible $n_{\rm shifts}$ range from 1 to 45, with 45 being the maximum due to the size of the kernel for this example (where $n_x=n_y=91$). Since the MCD-SOLA inversion, in some sense, utilizes all possible shifts once, the real-space method with 45 shifts and the multi-channel method should agree.

\begin{figure}
  \centering
  \includegraphics[width=\textwidth]{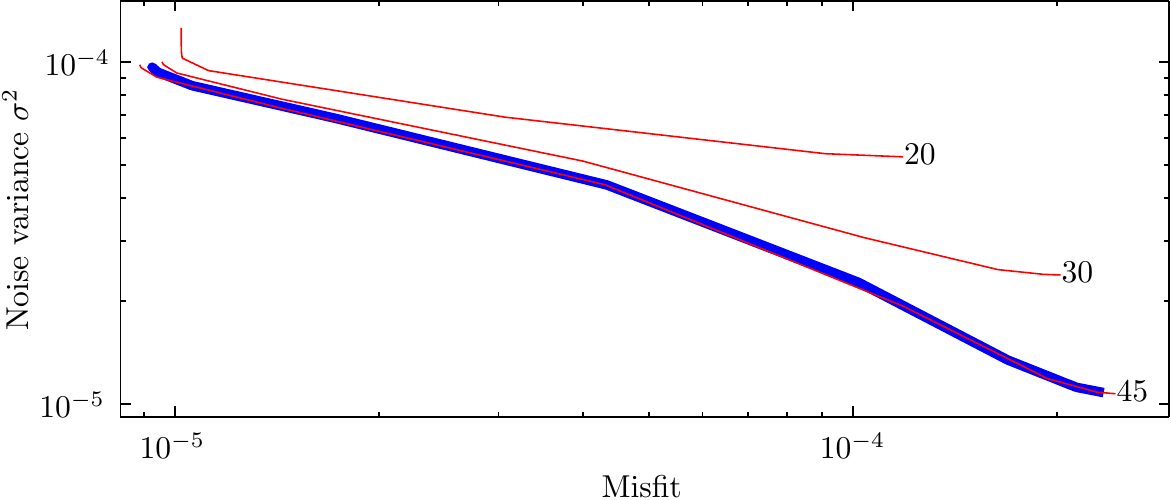}\vspace{.4cm}
  \includegraphics[width=\textwidth]{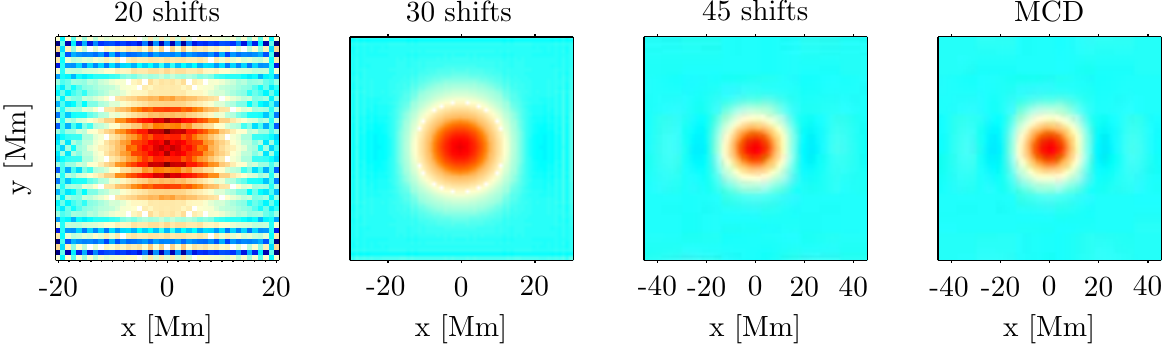}
  \caption{Example inversion using both the real-space and MCD methods and one single sensitivity kernel. The top panel shows trade-off curves where the noise unit is the fractional difference in sound speed [$\delta c^2/c^2$]. The thin red lines are for the real-space inversion using an increasing number of shifts from top to bottom indicated by the numbers. The thick blue line is the Fourier-space inversion.  The compute time for the 45-shift inversion was about a factor of 100 larger than for the MCD inversion.  The bottom panel corresponds to the inversion weights at trade-off parameter $\mu=0.01$~Mm$^{-3}$. This is the topmost point of each curve.  Note that the size of the weights in the real-space SOLA inversion depends on the number of shifts, which is why the spatial scale changes, although the color scale is fixed throughout.}
  \label{lcurve1}
\end{figure}

\begin{figure}
  \centering
  \includegraphics[width=\textwidth]{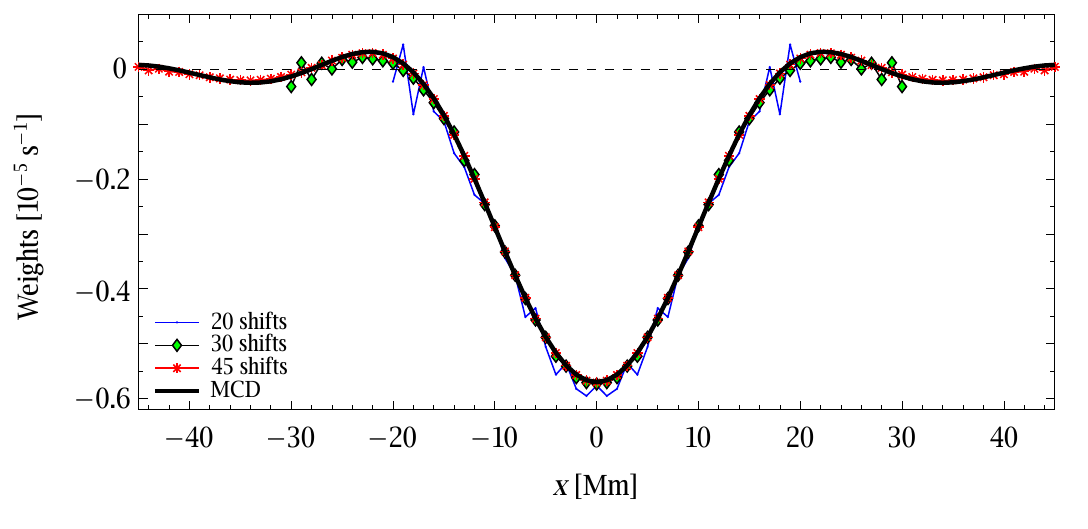}
  \caption{Cut through each inversion weight in Figure~\ref{lcurve1} along the line $y=0$. The curve for the 45-shift real-space inversion and the MCD-SOLA inversion are almost indistinguishable.}
  \label{cut}
\end{figure}

In Figure~\ref{lcurve1} we show the results for these inversions. The top panel of Figure~\ref{lcurve1} shows the trade-off curves, with red lines indicating the real-space inversion for varying shifts indicated by the numbers at the bottom of the curves. The thick blue line is the MCD inversion trade-off curve. These curves are typically plotted as the square of the random noise level versus the misfit on a logarithmic scale. We see that for increasing $n_{\rm shifts}$ the real-space inversion solution tends to the MCD solution. In fact, the L-curve for the 45-shift inversion falls on top of the one for the MCD inversion. Also shown in Figure~\ref{lcurve1} is a particular inversion weights $w_{\rm mn}^{c^2}$ for each inversion,  chosen from the first (topmost) point on each trade-off curve when $\mu=0.01$~Mm$^{-3}$. These are the points where the averaging kernel and target match best, \textit{i.e.} smallest misfit. This is a reasonable choice since our main concern here is not the noise,  which is still quite small anyway. The last two weights in Figure~\ref{lcurve1} demonstrate what the L-curves already suggest: the solutions of the two types of inversions are perfectly comparable when we take the maximum allowable number of shifts in the real-space method. The weights for inversions with a smaller number of shifts are quite ill-behaved due to edge effects, and in practice one actually never uses all possible shifts since it is computationally impractical to do so. This suggests that standard 3D-OLA inversions might have undesirable properties in the solution due to the necessary truncation of the problem. Cuts through all weights are shown in Figure~\ref{cut},  reinforcing this point when only a subset of shifts is used.

We recorded the computation time for each inversion. For the 45-shift case, the convolution matrix size is 8281 $\times$ 8281. The real-space inversion took  two orders of magnitude  longer to compute than the Fourier inversion (100 seconds compared to 1 second). This distinction only gets larger as the problem gets larger. Simply stated, the Fourier inversion takes a fraction of the time for small problems; for large problems, the real-space inversion is computationally intractable.

\begin{figure}
  \centerline{
    \includegraphics[width=.45\textwidth]{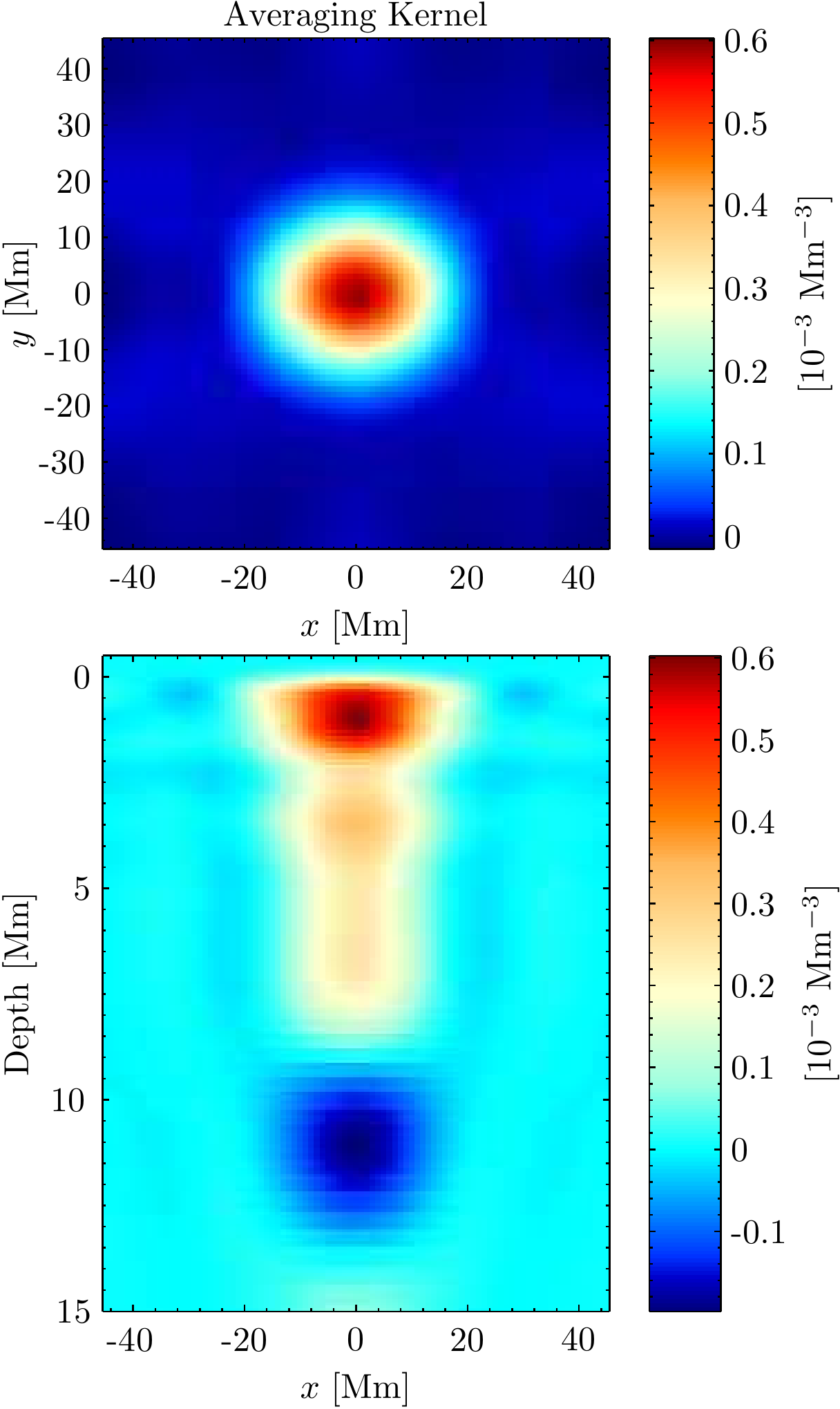}\hspace{.05\textwidth}
    \includegraphics[width=.45\textwidth]{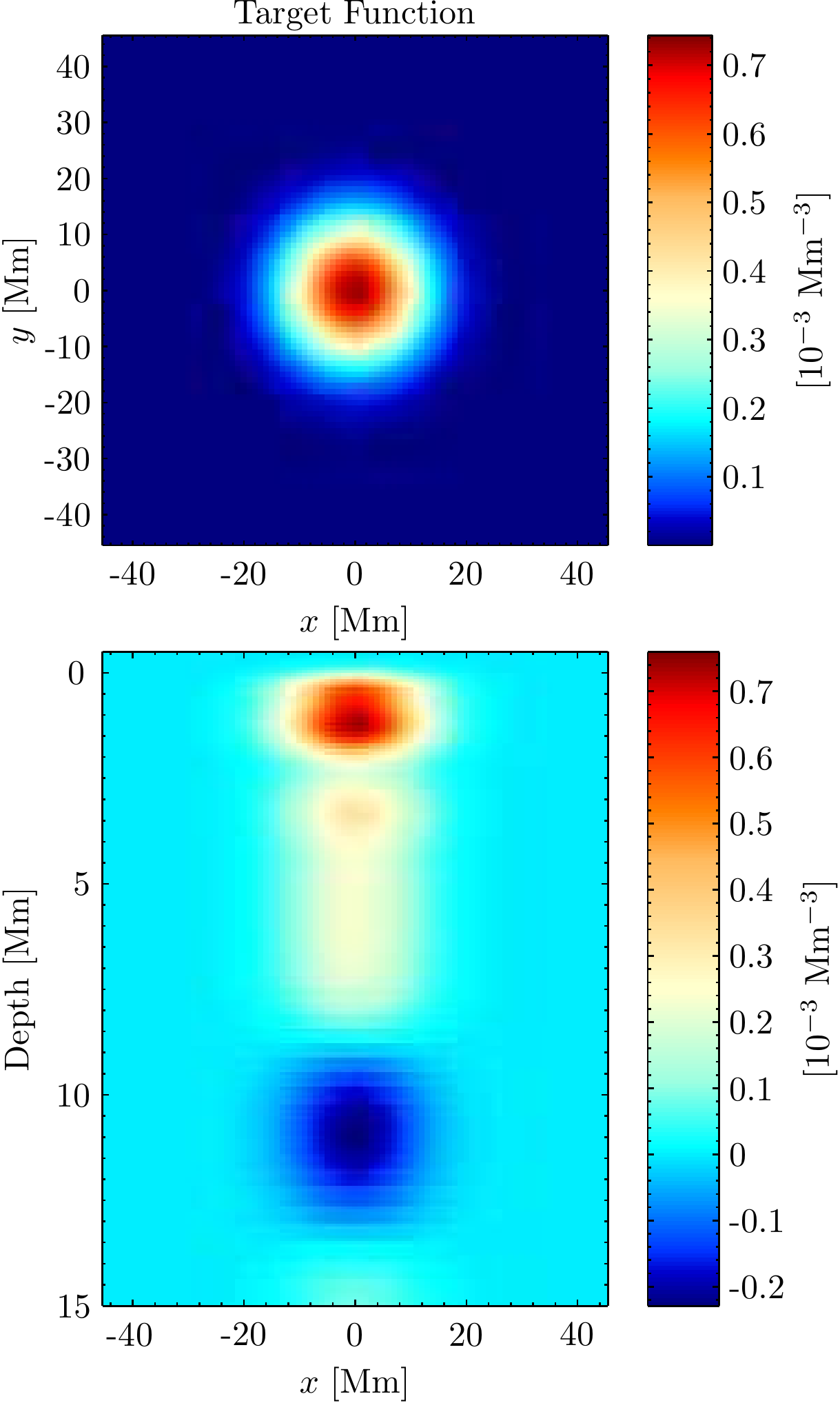}
  }
  \caption{Comparison of averaging kernel (left column) and target function (right column) from the example MCD-SOLA inversion. The target is the same one shown in Figure~\ref{senskern}. The top panels are slices through the averaging and target functions at a depth of 1 Mm. The bottom panels are slices with depth through the averaging and target functions along the $y=0$ line. }
  \label{fig:akern-targ}
\end{figure}

\begin{figure}
  \center
  \includegraphics[width=\textwidth]{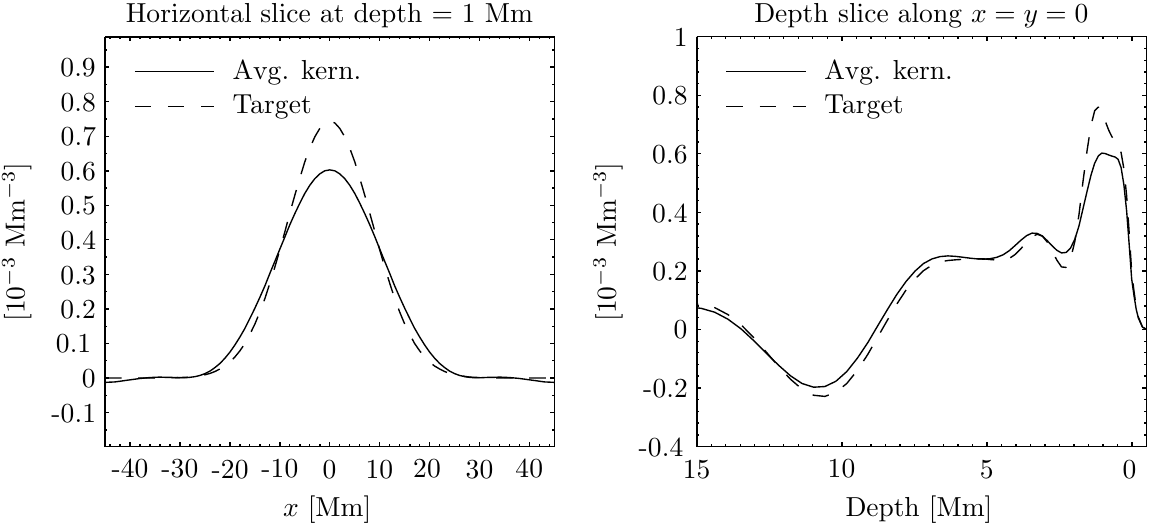}
  \caption{One-dimensional cuts through the  averaging kernel and target function from Figure~\ref{fig:akern-targ} for our example MCD inversion. The left panel is a horizontal slice through each function at a depth of 1 Mm. The right panel is a plot along the $x=y=0$ line with depth.}
  \label{avgkern}
\end{figure}

To show that the inversion does indeed  work, in Figures~\ref{fig:akern-targ} and \ref{avgkern} we provide  comparisons between averaging kernel and target from the MCD solution. In the horizontal  direction the agreement is quite acceptable, especially considering we have used only \textit{one} input sensitivity kernel. Since the vertical profile of the input target function was constructed to match that of the sensitivity kernel, the good agreement there is expected.

\section{Discussion and Conclusions}

The example considered here is a very simple, toy inversion to demonstrate the usefulness of this new Fourier-based MCD-SOLA method. We have also experimented with a larger problem whereby we consider point-to-point measurements of various orientations of the observation points with respect to the $x$-axis. We input the same kernel as the one shown in this work, as well as horizontal rotations of it to match the measurements. Using the MCD method, we found that only five rotations, spaced evenly between 0 and 90 degrees and keeping $\Delta$ fixed, are needed to find a very good averaging kernel that does not change with the addition of  more rotations. The computation time for this inversion was about 1.5 seconds. Had we attempted to solve the same problem with the real-space OLA inversion,  in addition to  consuming 40 gigabytes of memory, it would have taken weeks to compute.

We also solved our toy problem with several kernels of various distances [$\Delta$]. This allows one to obtain some resolving power in depth since the sensitivities differ. A target function was chosen as a 3D Gaussian peaked at a depth of 4 Mm beneath the surface. The MCD-SOLA inversion was able to find, as expected, an almost identical averaging kernel as the standard SOLA method.   For a realistic application of the MCD-SOLA method with various target depths from 5 Mm to the surface, we refer the reader to the recent work of  \citet{Svanda2011}, who inverted for vector flows using synthetic travel-time observations as input.

In conclusion, in this article we have extended what was originally done for RLS inversions  \citep{jacobsen1999} to a SOLA inversion. A toy example inversion problem was solved with this new approach to compare and contrast to the more standard real-space SOLA method. The example proved that the MCD-SOLA works completely satisfactorily while the real-space counterpart may be intractable for all but the smallest problems. In fact, we demonstrated that for a realistic helioseismic problem, the MCD-SOLA method can be orders of magnitude more computationally efficient than the corresponding real-space method.  We focused here on applications  to time--distance helioseismology, but this approach is completely generalizable to any local helioseismic method requiring inversions, such as ring diagram analysis and acoustic holography.  With the vast amounts of seismic data from the \textit{Solar Dynamics Observatory} (SDO), it is imperative to have efficient and consistent local helioseismic OLA inversion procedures for studying the solar interior.

\begin{acks}
This study was supported by the European Research Council under the  European  Community's Seventh Framework Programme (FP7/2007--2013)/ERC grant agreement  \#210949,  ``Seismic Imaging of the Solar Interior'', to PI L. Gizon (progress toward Milestone \#5). Additional support from the German Aerospace Center (DLR) through project ``German Data Center for SDO'' is acknowledged. ACB acknowledges support from NASA contracts NNH09CF68C, NNH07CD25C, and NNH09CE41C. M\v{S} acknowledges partial support from the Grant Agency of the Academy of Sciences of the Czech Republic under grant IAA30030808.
\end{acks}

%
%
\newcommand{\adv}{    {\it Adv. Spa. Res.}}
\newcommand{\annG}{   {\it Annales Geophysicae}}
\newcommand{\aap}{    {\it Astron. Astrophys.}}
\newcommand{\aaps}{   {\it Astron. Astrophys. Suppl.}}
\newcommand{\aapr}{   {\it Astron. Astrophys. Rev.}}
\newcommand{\ag}{     {\it Ann. Geophys.}}
\newcommand{\aj}{     {\it Astronom. J.}}
\newcommand{\apj}{{\it Astrophys. J.}}
\newcommand{\apjl}{{\it Astrophys. J. Lett.}}
\newcommand{\apjs}{{\it Astrophys. J. Suppl.}}
\newcommand{\apss}{   {\it Astrophys. Spa. Sci.}}
\newcommand{\cjaa}{   {\it Chinese J. Astron. Astrophys.}}
\newcommand{\gafd}{   {\it Geophys. Astrophys. Fluid Dyn.}}
\newcommand{\grl}{    {\it Geophys. Res. Lett.}}
\newcommand{\ijga}{   {\it Int. J. Geomag. Aeron.}}
\newcommand{\jastp}{  {\it J. Atmos. Sol. Terr. Phys.}}
\newcommand{\jgr}{    {\it J. Geophys. Res.}}
\newcommand{\mnras}{  {\it Mon. Not. Roy. Astron. Soc.}}
\newcommand{\physscr}{  {\it Physica Scripta}}
\newcommand{\nat}{    {\it Nature}}
\newcommand{\pasp}{   {\it Pub. Astron. Soc. Pac.}}
\newcommand{\pasj}{   {\it Pub. Astron. Soc. Japan}}
\newcommand{\pre}{    {\it Phys. Rev. E}}
\newcommand{\solphys}{{\it Solar Phys.}}
\newcommand{\sovast}{ {\it Sov. Astronom.}}
\newcommand{\ssr}{    {\it Space Sci. Rev.}}

\bibliographystyle{spr-mp-sola}                                                                                                                 
\bibliography{inversions}

\end{article}
\end{document}